# Stochastic TCO minimization for Video Transmission over IP Networks

Pejman Goudarzi

Multimedia systems group, IT faculty of Iran Telecom Research Center (ITRC)

End of the north Karegar st., Tehran-Iran

Email: pgoudarzi@itrc.ac.ir

*Abstract*—**From the viewpoint of service operators the Total Cost of Ownership (TCO) for developing a communication service comprises from two parts; CAPital EXpenditure (CAPEX) and OPerational EXpenditure (OPEX). These two types of costs are interrelated and affect any service provider's deployment strategy. In many traditional methods, selection of critical elements of a new service is performed in a heuristic manner aimed at reducing only the OPEX part of the TCO which is not necessarily optimal. Furthermore, exact cost modeling for such services is not always possible and contains some uncertainties. In the current work, after cost modeling of each video streaming element by capturing the effect of the model uncertainties, the TCO optimization problem for video streaming over IP networks is formulated as a stochastic optimization problem. The solution of the proposed optimization problem can cope with the cost modeling uncertainties and track the dynamism in the TCO and lead to a time-varying optimal solution. Numerical analysis results verify the developed method.**

## I. INTRODUCTION

Video streaming over IP networks is usually composed of some Video Content Delivery Network (VCDN) which is a system of computers containing copies of video data, placed at various points in a network so as to maximize bandwidth for access to the video data from clients throughout the network. A client accesses a copy of the video data near to the client, as opposed to all clients accessing the same central server, so as to avoid bottleneck near that server. VCDN nodes are usually deployed in multiple locations, often over multiple backbones. These nodes cooperate with each other to satisfy requests for video content by end users, transparently moving content to optimize the delivery process. Optimization can take the form of reducing bandwidth costs, improving end-user performance (reducing page load times and improving user experience), or increasing global availability of content.

The number of nodes and servers making up VCDN varies, depending on the architecture, some reaching thousands of nodes with tens of thousands of servers on many remote Point of Presences (PoPs). Others build a global network and have a small number of geographical PoPs.

Requests for video content are typically algorithmically directed to nodes that are optimal in some way. When optimizing for performance, locations that are best for serving content to the user may be chosen. This may be measured by choosing locations that are the fewest hops, the fewest number of network seconds away from the requesting client, or the highest availability in terms of server performance (both current and historical), so as to optimize video delivery across local networks. When optimizing for cost, locations that are least expensive may be chosen instead.

In an optimal scenario, these two goals tend to align, as servers that are close to the end user at the edge of the network may have an advantage in performance or cost. The Edge Network is grown outward from the origin/s by further acquiring (via purchase, peering, or exchange) co-locations facilities, bandwidth and servers.

Video streaming service[1] use IP as the transport platform to send video signals to the television via high-speed Internet connections such as fiber-to-the-X connections (FTTx) and/or next-generation digital subscriber lines (xDSL). With this technology, consumers will be in complete control of what, when, and where they watch television programming. Moreover, given the versatility of the IP network, consumers will have the opportunity to embrace a plethora of services that go beyond video signals [1].

Total cost of ownership (TCO) is a financial estimate designed to help consumers and enterprise managers assess direct and indirect costs commonly

---

[1] From now on and for the simplicity we replace in some places the term *video streaming service* with the word *service*.

related to software or hardware. It is a form of full cost accounting.

TCO modeling is a tool that systematically accounts for all costs related to an IT investment decision. TCO models were initially developed by Gartner Research Corporation in 1987 and are now widely accepted. Simply stated, TCO includes all costs, direct and indirect, incurred throughout the life cycle of an asset, including acquisition and procurement, operations and maintenance, and end-of-life management.

TCO analysis originated with the Gartner group in 1987 and has since been developed in a number of different methodologies and software tools. A TCO assessment ideally offers a final statement reflecting not only the cost of purchase but all aspects in the further use and maintenance of the equipment, device, or system considered. This includes the costs of training support personnel and the users of the system, costs associated with failure or outage (planned and unplanned), diminished performance incidents (i.e. if users are kept waiting), costs of security breaches (in loss of reputation and recovery costs), costs of disaster preparedness and recovery, floor space, electricity, development expenses, testing infrastructure and expenses, quality assurance, boot image control, marginal incremental growth, decommissioning, e-waste handling, and more.

TCO for developing a communication service comprises from two parts; CAPital EXpenditure (CAPEX) and OPerational EXpenditure (OPEX). These two types of costs are interrelated and affect any service provider's deployment strategy.

Middleware and other systems needed to provide video are also part of the total CAPEX. In a business case, CAPEX can be broken into fixed and variable parts; fixed being those costs to build the requisite system and infrastructure to deliver the services, and variable (usually unknown or uncertain) being those costs incurred with individual subscriber take rates. Consumer Premises Equipments (CPE) and in home installation are considered variable costs, along with DSL line cards, since the CAPEX is incurred only when service is taken. Ideally, fixed CAPEX should be minimized since it is the "at risk" investment to enter into the business. Variable CAPEX, although directly related to actual service take rate and revenue, cannot be so excessive as to present a Return on Investment (RoI) that it creates unacceptable ROI.

Internet Protocol TeleVision (IPTV) business cases as well as actual deployments have shown that the in home CPE and installation costs amount to 60% or more of the total installed cost for the IPTV system [2].

With CPE and in-home installation representing the largest portion of total installed cost, it is the area best targeted for cost reduction.

OPEX is composed of funds used by a company to acquire or upgrade physical assets such as property, industrial buildings or equipment. Like CAPEX, OPEX do contain some uncertain part which makes its exact mathematical modeling a complicated task.

Video-on-Demand (VoD) is the next killer application for video streaming and a subset of the IPTV service. Initial trials have been well received by customers and network operators are deploying VoD to increase subscriber revenues and service profitability. VoD allows subscribers to request the programming of their choice, when they want where they want it. It is this flexibility that appeals to the broader customer base when compared with regularly scheduled network programming of broadcast video.

In most traditional methods, the only objective is to minimize the OPEX part of the TCO by selecting critical components of the service in a heuristic manner. But, this approach may not necessarily result in optimal solution for the service providers. For example, in deploying the IPTV service in Iran, the service providers select the number of the required edge servers in order to minimize the OPEX part of the TCO [3].

Because of its static nature, this method doesn't consider the interrelations between OPEX and CAPEX which varies with time. For example, though choosing a specific initial number of edge servers may be optimal at the first stages of service deployment, this may not lead to an optimal solution for TCO minimization problem as time elapses.

Any solution for minimizing the TCO must take into the account the dynamic characteristics of the problem as time elapses.

In the current work, at first, bearing in mind the uncertain features of CAPEX and OPEX models, we try to develop an uncertain mathematical model for them. Then, based on the *stochastic optimization* techniques [4], a mathematical solution approach is developed to minimize the TCO. The proposed method tracks the dynamic changes in the number of subscribers and takes into the account the subscribers' geographical distributions and time. The rest of the paper is organized as follows:

In Section II, we have an overview on the related works. In Section III, a brief description of the video streaming service and its comprising elements is given. Section IV is about the TCO modeling and formulating the TCO minimization problem. At first the problem is formulated using appropriate models for CAPEX and OPEX parts of each video streaming element, then, using appropriate numerical methods inspired from the stochastic optimization theory, a cost optimization strategy is developed. Section V describes simulation results and we end the paper with some concluding remarks in the Section VI.

## II. RELATED WORKS

Although there exist few related works in this field but, we try to describe the available works and contrast the presented work with them.

In [5], the authors provide a design system for obtaining minimum cost survivable telecommunications networks. They integrate heuristics for obtaining survivable topologies and improving the cost of the network with heuristics for provisioning capacity. The proposed heuristics are based on the characteristic of the underlying graph.

The authors in [6] provide an overall methodology for combining hardware and network designs in a single cost minimization problem for multisite computer systems. Costs are minimized by applying a heuristic optimization approach to a sound decomposition of the problem. The authors consider most of the design alternatives that are enabled by current hardware and network technologies, including server sizing, localization of multi-tier applications, and reuse of legacy systems. The proposed methodology is empirically verified with a database of costs that has also been built as part of the paper.

The work presented in [7] deals with estimating performance measures, such as average response time for spatially distributed networks. The mentioned work, presents approximations which are tested using simulation and found to give good results.

In [8], which is the most similar one to the current work, the TCO optimization problem of IPTV service is formulated as a nonlinear programming one. The solution of the proposed optimization problem can track the dynamic changes of the TCO and lead to a time-varying optimal solution.

Although as in [5-6] the objective in this work is finding a minimum cost network, but, the presented work differs from [5-6] in the fact that, in this work a stochastic optimization techniques are employed which can capture more exactly the uncertainties in the system cost modeling than the heuristics adopted in [5-6].

The work also is different from [7] in estimating different performance measure parameters of the networks which in the case of the current paper is the networks' TCO.

The presented work, differs from the work presented in [3] in the fact that in [3], the cost modeling is performed by incorporating some specific assumptions about the time and scale evolution of CAPEX and OPEX of each IPTV component. In the current work, by incorporating some uncertainty parameters in cost, we will present more realistic models for these costs. The cost models in the presented work may have multiple optima. The position of these optimal points may even change in time and population in a random manner. Hence, by introducing the concept of *stochastic optimization* theory we try to converge to these optimal solutions.

The presented work, also differs from the work presented in [8] in the fact that in [8], the cost modeling is performed by incorporating some specific assumptions about the time and scale evolution of CAPEX and OPEX of each IPTV component. But, in the current work, by incorporating some uncertainty parameters in cost, we will present more realistic models for these costs. The cost models in the presented work may have multiple optima. The position of these optimal points may even change in time and population in a random manner. Hence, by introducing the concept of *stochastic optimization* theory we try to converge to these optimal solutions.

## III. VIDEO STREAMING SERVICE

Video streaming service describes a system where a digital video service is delivered using the Internet protocol over a network infrastructure, which may include delivery by a broadband connection [9].

For residential users, video streaming is often provided in conjunction with VoD or IPTV and may be bundled with Internet services such as Web access and VoIP.

This service is typically supplied by a broadband operator using a closed network infrastructure. This closed network approach is in competition with the delivery of video content over the public Internet.

In businesses, video streaming may be used to deliver video content over corporate LANs and business networks.

A typical video streaming service delivery scenario is depicted in the Fig.1.

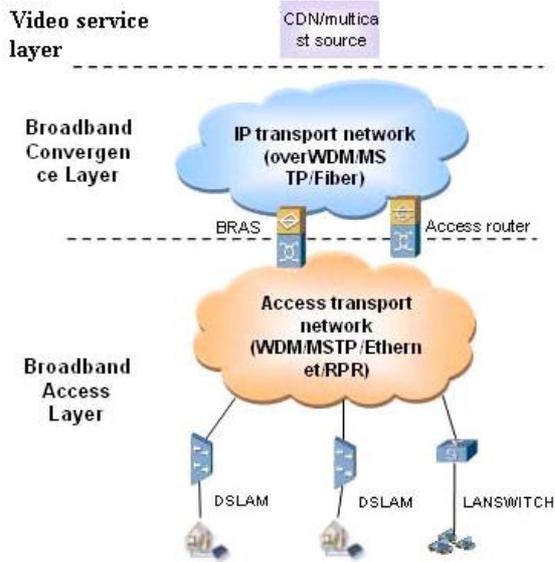

**Figure 1.** Typical video streaming over IP service scenario

The basic components of video delivery service are, video streaming servers [10], edge streaming servers used for load balancing purposes, encoded content, transport and access Quality of Service (QoS)-enabled networks [11], BRAS (Broadband Remote Access Server), DSLAM (Digital Subscriber Line Access Multiplexer), STB (Set Top Box) and ADSL (Asymmetric DSL) modems.

Each component is associated with an incurred CAPEX and OPEX. Some components such as transport network are out of the service provider's control and impose only a long-term OPEX on the service provider's deployment strategy and some of the components such as content only consist of an initial CAPEX and don't impose any important OPEX on the service development.

In the following sections, mathematical models for the CAPEX and OPEX of each component are developed and based on the proposed models; a dynamic solution for the TCO minimization problem is introduced.

## IV. PROBLEM FORMULATION

In order to minimize the TCO (CAPEX+OPEX) of service, we have developed in a heuristic manner, the mathematical models associated with the CAPEX and OPEX of each component which is involved in the service.

In practice, development of a comprehensive model for minimizing the total cost of ownership of a service is very difficult. In this section we ignore some parameters described in previous section and adopt a similar approach such as [12] to make the mathematical modeling possible.

DSLAM, STB, ADSL modem, Edge Server, Main server are among the devices which together build a video streaming solution. In this section, mathematical models for CAPEX and OPEX of these components are presented. To do so, we first make an estimation of the initial cost of each component individually based on the research accomplished in [2]. The actual costs may be composed of these initial costs augmented by some unknown perturbations. The initial cost of STB, ADSL modem, DSLAM, BRAS, main server, edge server, content and infrastructure are denoted by $x_{STB}$, $x_{MDM}$, $x_{DSLM}$, $x_{BRAS}$, $x_{MSRVR}$, $x_{ESRVR}$, $x_{CNT}$, $x_{INSTR}$ respectively.

Table 1 shows the initial cost of the devices normalized by the initial cost of ADSL modem.

TABLE 1. INITIAL COST OF IPTV DEVICES NORMALIZED BY ADSL MODEM COST

| DEVICE | COST | DEVICE | COST |
|---|---|---|---|
| $x_{STB}$ | 3.33 | $x_{MSRVR}$ | 176 |
| $x_{MDM}$ | 1 | $x_{ESRVR}$ | 44.44 |
| $x_{DSLM}$ | 128 | $x_{CNT}$ | 11.11 |
| $x_{BRAS}$ | 10 | $x_{INSTR}$ | 100 |

It must be mentioned that the above normalized values in table (1) are obtained from a survey on the prices of each service component which was investigated about the Iran's IT market in [2].

In order to cover the dynamic nature of the mentioned components of the TCO, the mathematical models for CAPEX and OPEX are chosen to be functions of time $t$, number of subscribers $n$ and number of edge servers $m$.

It must be mentioned that except the CAPEX associated with the content (which assumed to be fix) and the CAPEX associated with the network infrastructure (which assumed to be null), all of the CAPEX formulations are in accordance with the power-law/exponential distributions for the cost estimation method and the inflation effect [13].

Hence, the CAPEX associated with most of the video streaming components is a decreasing function of the number of subscribers $n$ and an increasing function of time $t$ because it is assumed that the CAPEX can be reduced as the demand $n$ increases and can be increased for the sake of inflation as time $t$ elapses.

It is assumed that there exist 2 main servers (original+backup) for the sake of high availability purposes [14].

The OPEX of each video streaming component is assumed initially to be zero and can increase if time elapses and/or the number of subscribers increases.

From research adopted in Iran, the OPEX associated with infrastructure part of the network is assumed to be a decreasing function of $m$ and an increasing function of both $n$ and $t$.

In the following equations, parameters beginning with 'C' represent CAPEX and those beginning with 'O' represent OPEX. In all of the following cost formulations, we denote the unknown perturbation parameters vector by $\bar{\xi}$. This uncertainty vector belongs to a compact uncertainty set known as $\Psi$. These perturbation parameters are some random variables which are added to compensate for the probably inexact cost models used for the TCO modeling of the service.

The perturbation parameter for element $i$ is considered to have a Gaussian (normal) distribution with zero-mean and standard deviation equal to square root of the initial cost $x$ as follows:

$$\xi_i \sim N(0, \sqrt{x_i}), \forall i$$

The CAPEX and OPEX are modeled as follows:

$$C_{STB}(n,t) = \left(x_{STB} + (y_{STB} - x_{STB})e^{-z_{STB}n}\right) \cdot \left(2 - e^{-e_0 t}\right) + \xi_{c,STB} \quad (1)$$
$$O_{STB}(n,t) = x_{STB}\left(1 - e^{-e_1 n}\right)\left(1 - e^{-e_0 t}\right) + \xi_{o,STB}$$

$$C_{MDM}(n,t) = \left(x_{MDM} + (y_{MDM} - x_{MDM})e^{-z_{MDM}n}\right) \cdot \left(2 - e^{-e_0 t}\right) + \xi_{c,MDM} \quad (2)$$
$$O_{MDM}(n,t) = x_{MDM}\left(1 - e^{-e_1 n}\right)\left(1 - e^{-e_0 t}\right) + \xi_{o,MDM}$$

The Eqs. 1 and 2 describe that without considering the perturbation term, CAPEX of the Set Top Box and ADSL modems are increasing functions of time $t$ and decreasing functions of the number of subscribers $n$ [13].

Furthermore, it is trivial that still without considering the perturbation term, the OPEX of these devices must be increasing functions of both time $t$ and number of subscribers $n$.

$$C_{DSLM}(t) = x_{DSLM} \cdot \left(2 - e^{-e_0 t}\right) + \xi_{c,DSLM} \quad (3)$$
$$O_{DSLM}(n,t) = x_{DSLM}\left(1 - e^{-e_1 n}\right)\left(1 - e^{-e_0 t}\right) + \xi_{o,DSLM}$$

$$C_{BRAS}(t) = x_{BRAS} \cdot \left(2 - e^{-e_0 t}\right) + \xi_{c,BRAS} \quad (4)$$
$$O_{BRAS}(n,t) = x_{BRAS}\left(1 - e^{-e_1 n}\right)\left(1 - e^{-e_0 t}\right) + \xi_{o,BRAS}$$

The Eqs. 3 and 4 describe that without considering the perturbation term, CAPEX of the DSLAM and BRAS are increasing functions of time $t$ but do not depend on the number of subscribers $n$ because the DSLAM or BRAS is bought only by the service providers and large companies not real persons, so their initial price (CAPEX) seems not to be dependent on the number of subscribers.

Furthermore, still without considering the perturbation term, the OPEX of these devices must be increasing functions of both time $t$ and number of subscribers $n$.

As Eq. 5 describes, without considering the perturbation term, the CAPEX of the media server is an increasing function of time $t$ but do not depend on the number of subscribers $n$ because there exist only two (original+backup) media servers in the video streaming development plan.

The OPEX of media server must be an increasing function of $t$ and $n$.

$$C_{MSRVR}(t) = x_{MSRVR} \cdot \left(2 - e^{-e_0 t}\right) + \xi_{c,MSRVR} \quad (5)$$
$$O_{MSRVR}(n,t) = x_{MSRVR}\left(1 - e^{-e_0 t}\right) + \xi_{o,MSRVR}$$

In Eq. 6, without considering the perturbation term, the CAPEX of the edge server is an increasing function of time $t$ but do not depend on the number of subscribers $n$ because it is just for the service providers' usage.

$$C_{ESRVR}(t) = x_{ESRVR} \cdot \left(2 - e^{-e_0 t}\right) + \xi_{c,ESRVR} \quad (6)$$
$$O_{ESRVR}(n,t) = x_{ESRVR}\left(1 - e^{-e_1 n}\right)\left(1 - e^{-e_0 t}\right) + \xi_{o,ESRVR}$$

The CAPEX associated with the content is assumed to be a fixed term plus perturbation term. Also it is assumed that there is not any OPEX associated with the contents (Eq.7).

$$C_{CNT} = x_{CNT} + \xi_{CNT} \quad (7)$$

It is assumed that the network infrastructure exists during the service development, so, there is not any CAPEX associated with this component.

As can be verified in Eq.8, with neglecting the perturbation effect, the OPEX of the network infrastructure is assumed to be increasing functions of $t$ and $n$ but decreasing function of the number of edge servers $m$ because of the load balancing feature which arise with increasing the number of edge servers [15].

$$O_{INSTR} = \left(x_{INSTR} + y_{INSTR}e^{-z_{INSTR}m}\right) \cdot \left(p_{INSTR} - q_{INSTR}e^{-s_{INSTR}n}\right) \cdot \left(v_{INSTR} - w_{INSTR}e^{-u_{INSTR}t}\right) + \xi_{o,INSTR} \quad (8)$$

Other parameters in Eqs. 1-9 are obtained based on the research conducted in Iran's IT market [2]:

$$y_{STB} = 0.8x_{STB}; \; y_{MDM} = 0.8x_{MDM}$$
$$y_{INSTR} = 0.11x_{INSTR}; \; z_{STB} = z_{MDM} = 3 \times 10^{-6};$$
$$z_{INSTR} = 0.002231; u_{INSTR} = 0.00886$$
$$p_{INSTR} = q_{INSTR} = s_{INSTR} = 1$$
$$v_{INSTR} = 1.1; \; w_{INSTR} = 0.1$$

The value of $e_0$ and $e_1$ for each device is considered as $0.1\ x$.

As we see in the above equations, the CAPEX of each component is a decreasing function of the number of subscribers $n$ and an increasing function of time $t$ because it is assumed that the CAPEX can be reduced as the request $n$ increases and can be increased for the sake of inflation as time $t$ evolves.

The OPEX of each component is assumed initially to be null and can increase as time and number of subscribers increase.

From research adopted in [3], the OPEX associated with infrastructure part of the network is assumed to be a decreasing function of $m$ and an increasing function of both $n$ and $t$.

Finally, TCO for the service is:

$$\begin{aligned}\text{TCO}(n,m,t,\overline{\xi}) = &n(C_{STB}+O_{STB})+n(C_{MDM}+O_{MDM})+ \\ &\left\lceil\frac{n}{\tau}\right\rceil\cdot(C_{DSLM}+O_{DSLM}+C_{BRAS}+O_{BRAS})+C_{CNT}+ \\ &2(C_{MSRVR}+O_{MSRVR})+m(C_{ESRVR}+O_{ESRVR})+O_{INSTR} \\ &n,m,t>0 \text{ and } \overline{\xi}\in\Psi\end{aligned} \quad (9)$$

Where, $\lceil x \rceil$ is the smallest integer number which is greater than or equal to $x$ and $\tau$ is the number of DSLAM ports. We have assumed that $\tau = 128$. As described earlier, one additional media server is considered for redundancy purposes.

In the current study, we have tried to track the changes which are imposed on the optimal TCO as time evolves by regulating the number of edge servers $m$ as a function of time. By research conducted in Iran's IT market, it is concluded that the number of edge servers can be considered as a major influencing parameter affecting the TCO of the video streaming solution.

Thus, based on Eq. 9, the objective is to choose the optimal value of $m$ which leads to a minimized TCO.

It is obvious and legal that the number of edge servers $m$ may not exceed the number of subscribers $n$, so we are faced with a constrained nonlinear and stochastic optimization problem [16].

In contrast to classical deterministic optimization which assumes that perfect information is available about the objective function (and derivatives, if relevant) and that this information is used to determine the search direction in a deterministic manner at every step of the algorithm [17], in many practical problems such as the current TCO minimization problem, due to the existence of the uncertain parameters, such exact information is not available [4].

In the optimization process, we assume that the objective function can be decoupled to the two specific terms as follows:

$$\text{TCO}(m) = L(m) + \tilde{N}(m) \quad (10)$$

Where, $L(.)$ and $\tilde{N}(.)$ are the exact loss function and noise terms respectively according to the notations presented in [4].

Note that the noise terms show dependence on $m$. This dependence is relevant for many applications. It indicates that the common statistical assumption of independent, identically distributed (i.i.d.) noise does not necessarily apply since $m$ will be changing as the search process proceeds.

Noise term, fundamentally alters the search and optimization process because the algorithm is getting potentially misleading information throughout the search process.

Based on the dependency between the noise and loss terms in Eq.10 and inadequate gradient information about the objective function TCO, we have selected the *stochastic approximation* (SA) method for finding the optimal point [18].

The recursive procedure of interest is in the general SA form as follows:

$$\hat{m}_{k+1} = \hat{m}_k - a_k \hat{g}_k(\hat{m}_k) \quad (11)$$

Where, $\hat{g}_k$ is the estimate of the true gradient $g$ at iteration $k$ of TCO in the Eq.10 and $a_k$ is some positive constant.

For finding this estimate, we have used the two-sided Finite Difference SA (FDSA) method as follows [18]:

$$\hat{g}_k(\hat{m}_k) = \frac{\text{TCO}(\hat{m}_k+c_k)-\text{TCO}(\hat{m}_k-c_k)}{2c_k} \quad (12)$$

Where, $c_k$ is some positive constant.

The choice of these gain sequences $a_k$ and $c_k$ is critical to the performance of the method. Common forms for the sequences are:

$$a_k = \frac{a}{(k+1+A)^\alpha}, \quad c_k = \frac{c}{(k+1)^\gamma}$$

Where the coefficients $a, c, \alpha,$ and $\gamma$ are strictly positive and $A \geq 0$. The user must choose these coefficients, a process usually based on a combination of the theoretical restrictions above, trial-and-error numerical experimentation, and basic problem knowledge. In some cases, it is possible to partially automate the selection of the gains.

It is shown in [19] that under above conditions $\hat{m}_k$ will eventually converge to the optimal $m^*$ in some appropriate stochastic sense such as *almost surely* (a.s.) [4].

In the current work, we have tried to track the changes which are imposed on the optimal TCO as time evolves by regulating the number of edge servers $m$ as a function of time. By work presented in [3] it is resulted that the number of edge servers can be regarded as a major parameter affecting the TCO of the solution.

This method captures the time variations and leads to the time-varying minimized TCO [8].

## V. NUMERICAL ANALYSIS

We have used the scenario depicted in the Fig.2 which is similar to one adopted in [8] for deployment of the service in Iran. In order to minimize the total cost (TCO) of the service in Eq.9, the minimization was performed with respect to *m* (number of edge servers) based on the Eq.11.

In other words, in Eq. 9, we have to determine the optimal value of *m* so that the total cost in this equation is minimized for a specific number of subscribers *n* and time *t*. Time means the number of years which have elapsed since service was first deployed (*t*=0).

In the scenario, it is assumed that the end-users are distributed uniformly in a circular geographical region and for each $\lceil n/\tau \rceil$ user there exist a BRAS and a DSLAM. Each edge server can be connected to at least one DSLAM for streaming more demanding contents.

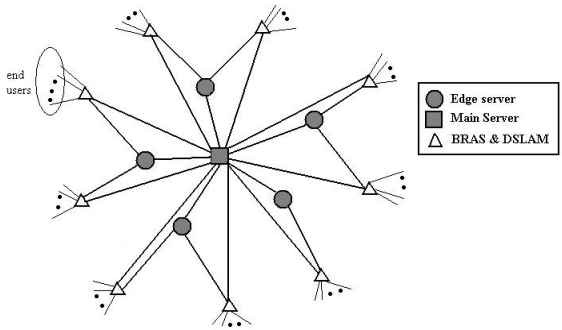

**Figure 2.** Simulated service scenario

As can be verified in the Fig.3, the TCO evolves in time according to the depicted figure. A time period of 50 years is considered for simulation purpose. This time variation is a function of various CAPEX and OPEX parameters and their associated perturbations.

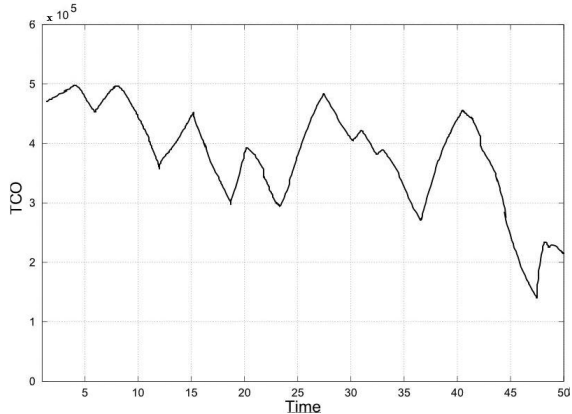

**Figure 3.** A sketch of the TCO vs. time (yrs)

In Fig.4, the dynamism of the TCO has been depicted versus the number of edge servers for different population sizes (*n*). In this figure, TCO has depicted versus the number of edge servers for *n*=5,000 (bold), *n*=10,000 (dashed) and *n*=50,000 (dotted). It can be deduced from this figure that by increasing the population size (*n*), the TCO has been raised.

It can also be verified in this figure that, because of the stochastic nature of the TCO components, the depicted TCOs may have multiple optima. So, traditional optimization tools such as those used in [16-17] cannot be used for finding the global optimal point. Because of the random nature of the TCO components, the stochastic optimization tools (Eqs.11-12) have been used in this scenario for finding the optimal solution.

The simulation parameters used in this work are listed in the Table 2. Other simulation parameters are selected based on the Table 1.

TABLE 2. SIMULATION PARAMETERS

| Parameter | Value | Parameter | Value |
|---|---|---|---|
| *a* | 10 | *γ* | 2 |
| *c* | 1 | *A* | 4 |
| *α* | 3 | | |

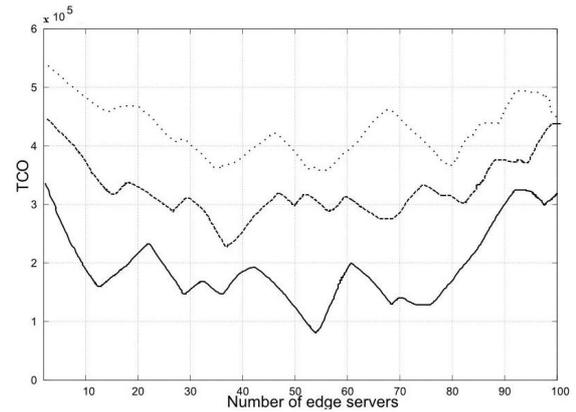

**Figure 4.** TCO vs. number of edge servers for *n*=5,000 (bold), *n*=10,000 (dashed) and *n*=50,000 (dotted)

In Fig. 5, it can be verified that optimal number of edge servers evolve in time (dashed line) such that the TCO in Eq. 9, goes to the global (and time-varying) minimum. The average number of edge servers in 5-year intervals is derived and depicted in this figure (bold-line).

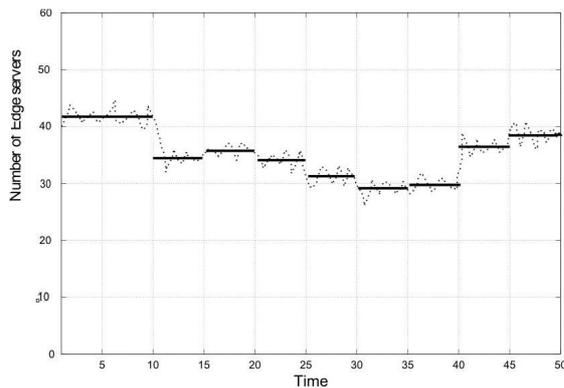

**Figure 5.** Time evolution of optimal (dashed) and average (solid) number of edge servers

## VI. CONCLUSION

Though video streaming (IPTV and VOD) services are very attractive to the end users, they are highly expensive and this has caused them to be developed with a very slow pace. Minimizing the TCO (CAPEX+ OPEX) of these services is a challenge to all video streaming service providers. In a video streaming scenario one of the major factors that determine the cost of the service is the number of edge servers. We used stochastic optimization technique for capturing the effect of the model uncertainty and minimizing the TCO using selection of an optimal number of edge servers in a typical video streaming scenario. This leads to the lowest cost of service deployment. The proposed algorithm has proved to be quite efficient and dynamic in minimizing the TCO of video streaming service as the number of subscribers increases and time elapses.

REFERENCES

[1] Zhone technologies, "In-home triple play delivery", white paper, 2004, USA.
[2] Thomson, J., "IPTV-market, regulatory trends and policy options in Europe", in proceedings of *IET IPTV Deployment and Service Delivery*, 2007.
[3] Goudarzi P., "Dynamic Total Cost of Ownership Optimization for IPTV Service Deployment", Journal of Applied Sciences, Vol. 9, No.4, pp. 707-715, 2009.
[4] Spall, J. C. (2003). Introduction to Stochastic Search and Optimization, Wiley.
[5] Clarke, L.W. and Anandalingam, G., "An integrated system for designing minimum cost survivable telecommunication network", *IEEE Trans. Syst. Man Cybern*, pp.856-862, 1996.
[6] Ardagna, D. Francalanci and C. Trubian, M., "Joint optimization of hardware and network costs for distributed computer systems", *IEEE Trans. Syst. Man Cybern*, pp.470-484, 2008.
[7] Berman, O. and Vasudeva S., "Approximating performance measure for public services", *IEEE Trans Syst. Man Cybern.*, pp.583-591, 2005.
[8] Goudarzi P., "Dynamic Total Cost of Ownership Optimization for IPTV Service Deployment", Journal of Applied Sciences, Vol. 9, No.4, pp. 707-715, 2009.
[9] ATIS IIF's IPTV Architecture Requirements, ATIS 0800002.
[10] Ardanga, D. and Chiara F., "A cost-oriented for the design of web based IT architectures", in proceedings of *ACM symposium on Applied Computing*, 2002.
[11] Abdelzaher, T.F., Shin, K.G. and Bhatti, N., "User-level QoS-adaptive resource management in server end-systems" IEEE Trans. Comput., pp.678-685, 2003.
[12] Heigden, F., Duin R., Ridder D. and Tax. D., *Classification parameter estimation and state estimation*, Wiley, 2004.
[13] Akoi, M. and Yoshikawa H., "Stock prices and the real economy: Power law versus exponential distributions", *Journal of Econ. Interact. Coordinat.,* pp.45-73, 2006.
[14] Sadowsky G., Dempsey J., Greenberg A. and Mack B., *IT security handbook*, 2003.
[15] Tanenbaum A.S., *Computer networks*, Prentice hall, 2002.
[16] Bertsekas, D. P., *Nonlinear Programming*: 2nd Edition. Athena Scientific, 1999.
[17] Luenberger, D.G., Linear and Nonlinear Programming, 2nd Ed. Addison-Wesley Publishing Company, 1984.
[18] Robbins H. and Monro S., "A Stochastic Approximation Method", *Annals of Mathematical Statistics*, pp. 400–407, 1991.
[19] Fabian, V. (1971), "Stochastic Approximation," in Optimizing Methods in Statistics (J.S. Rustigi, ed.), Academic Press, New York, pp. 439–470.